\documentstyle[aps]{revtex}

\begin{document}
\title{Determination of the $S_{18}$ astrophysical factor for $^{8}$B($p$,$\gamma $)%
$^{9}$C from the breakup of $^{9}$C at intermediate energies}
\author{L. Trache$^{1}$, F. Carstoiu$^{2,3}$, A. M. Mukhamedzhanov$^{1}$ and R. E.
Tribble$^{1}$}
\address{$^{1}$Cyclotron Institute, Texas A\&M University, College Station, TX
77843-3366, USA \\
$^{2}$Institute of Physics and Nuclear Engineering H. Hulubei, Bucharest,
Romania\\
$^{3}$Laboratoire de Physique Corpusculaire, F-14050 Caen Cedex, France}
\date{\today}
\maketitle

\begin{abstract}
We have used existing data on the one-proton-removal cross section of $^{9}$%
C at 285 MeV/u and Glauber model calculations to extract the asymptotic
normalization coefficient for the wave function of the last proton in the
ground state of $^{9}$C. The calculations are done first using folded
potentials starting from two different effective nucleon-nucleon
interactions and second in the optical limit using three nucleon-nucleon
interactions, and the results are found to be consistent, with no new
parameters adjusted. We find $C^{2}(p_{3/2})+C^{2}(p_{1/2})=1.22\pm 0.13$ fm$%
^{-1}$. From this result we obtain the astrophysical factor for the proton
radiative capture reaction $^{8}$B($p$,$\gamma $)$^{9}$C as $S_{18}(0)=46\pm
6$ eV$\cdot $b. The calculated energy dependence of the astrophysical
S-factor for the energy region $E_{cm}=0-0.8$ MeV and the reaction rates for 
$T_{9}=0-1\ $are included.
\end{abstract}


\section{Introduction}

Radiative capture reactions are of crucial importance in nuclear
astrophysics. The capture of charged particles on nuclei, in particular of
protons in basic processes like hydrogen burning in various stellar
environments, is very much hindered by the Coulomb repulsion. This leads to
very small cross sections and consequently to the well known experimental
problems associated with direct determination of astrophysical $S$-factors
at very low energies \cite{rolfs}. Moreover, with the large number of
reaction chains found to be of importance in nucleosynthesis calculations
for different static and explosive burning scenarios \cite
{wies89,cha92,kapp98}, more data involving the capture on unstable nuclei
becomes necessary. In many instances direct measurements involving unstable
nuclei are very difficult or even impossible and indirect methods must be
used.

It has been known for a long time that transfer reactions can be used as
indirect methods for nuclear astrophysics (see, e.g., \cite{rolfs72,rolfs}
and references therein) and various techniques have been used since. A few
years ago another indirect approach based on measurements of peripheral
transfer reactions was proposed \cite{akr90,xu}, and was subsequently used
to determine the astrophysical factor $S_{17}(0)$ for the reaction $^{7}$%
Be(p,$\gamma $)$^{8}$B using ($^{7}$Be,$^{8}$B) transfer reactions. The
method, which is based on the observation that radiative capture of protons
is a very peripheral process at astrophysical energies, involves the
extraction of nuclear quantities called Asymptotic Normalization
Coefficients (ANC) from proton transfer reactions and these ANCs are then
used to determine the astrophysical cross sections. The method works if the
transfer reactions, which have much higher cross sections than the capture
reactions, are also peripheral. Using secondary beams available today, such
experiments can be done in a matter of days \cite
{azhari1,azhari3,liu,beaumel,tang}. More recently we have shown \cite{tra01}
that one-nucleon-removal reactions offer an alternative and complementary
technique for extracting ANCs which is particularly well adapted to rare
isotopes produced using fragmentation. In the breakup of loosely bound
nuclei at intermediate energies, the requirement of core survival in the
final channel automatically selects core-target interactions which are
highly peripheral, with the result that the wave function of the removed
nucleon is probed at and beyond the core's nuclear surface. This is then
very similar to the low-energy light ion transfer reactions where the short
mean free path of the ions in the optical potential leads to surface
localization. In Ref. \cite{tra01} $^{8}$B breakup data at energies between
30 and 300 MeV/u on a variety of targets were compared with calculations
done using a Glauber model \cite{Glauber} to extract the same ANC as that
obtained from transfer reactions at about 12 MeV/u. We suggested that such
experiments have the additional advantage of using beams of lesser quality,
at higher energies, and therefore can be applied to nuclei farther away from
the stability.

Here we use a similar approach to analyze the breakup of $^{9}$C. Existing
cross section data for the one-proton-removal reaction with $^{9}$C at 285
MeV/u on four targets \cite{blank} are used to extract the ANC for the
virtual process $^{9}$C$\rightarrow ^{8}$B$+p$. The breakup cross sections
are calculated with a Glauber type model, first using the potential approach
described in \cite{tra01} now with two types of effective interactions, and
then using a Glauber model in the optical limit, with three different
prescriptions for the elementary nucleon-nucleon (NN) scattering amplitudes.
Good consistency is obtained between the different approaches. The value of
the ANC is then used to determine the astrophysical factor $S_{18}$ for the
proton radiative capture reaction $^{8}$B($p,\gamma $)$^{9}$C in the energy
range $E_{cm}=0-0.8$ MeV and the reaction rate in stellar environments is
obtained for the temperature range $T_{9}=0-1$. The ANC extracted here
agrees with that obtained recently from a transfer reaction d($^{8}$B,$^{9}$%
C)n on a deuteron target at 14.4 MeV/u \cite{beaumel}, but our result has a
smaller uncertainty.

\section{The breakup of $^{9}$C}

Much effort has been devoted in recent years to the study of exotic nuclei.
In particular, it has been shown that one-nucleon removal reactions with
radioactive projectiles at intermediate energies can be used to study their
structure \cite{tanihata96,hansen01,cortina,tostev}. Typically an exotic
nucleus $B=(Ap)$, where $B$ is a bound state of the core $A$ and the nucleon 
$p$, is produced by fragmentation from a primary beam, separated and then
used to bombard a secondary target. After breakup occurs, the cross section
and parallel momentum distribution of the core $A$ are measured. It has been
shown that the parallel momentum distribution of the detected core $A$ can
be used to establish the orbital momentum of the relative motion of the
nucleon, even disentangle contributions from different contributing
orbitals, and that coincidences with gamma-rays enables one to separate the
contribution of different core states into the ground state of the
projectile. Typically the integrated cross sections have been used to
extract absolute spectroscopic factors. However, we have shown \cite
{tra01,tra02} that the extracted spectroscopic factors depend strongly on
the assumed single particle wave function (or equivalently on the geometry
of the binding two-body potential that produces it), and unfortunately, for
exotic nuclei these potentials are poorly known and there is not much other
information to put constraints on them. We have shown that because these
breakup processes are essentially peripheral, especially for loosely bound
nuclei, one can obtain the asymptotic normalization coefficients (ANCs),
rather than the spectroscopic factors. The ANC only gives information about
the asymptotic tail of the wave function for the last nucleon, but does not
say anything about the behavior of the many body wave function inside the
nucleus. Fortunately, this is all we need to determine peripheral
observables, chief among them the astrophysical $S$-factor for radiative
proton capture reactions. The extraction of ANC also has the advantage of
not depending on the fundamental assumption about the validity of the
independent particle model in the whole nucleus, but only at its surface
(see \cite{pandar,brown} and the references therein, for a discussion on
this subject), or of the cluster-like structure of some loosely bound
nuclei, an important characteristic.

We apply here the method described first in \cite{tra01}, to analyze
existing one-proton-removal cross section data for $^{9}$C at 285 MeV/u on
four different targets (C, Al, Sn and Pb) \cite{blank}. In these experiments
a beam of radioactive $^{9}$C nuclei stroke the target, and the residual $%
^{8}$B was detected. We use the same extended Glauber model \cite
{nego,sauvan,carst-glaub} to calculate the cross sections. The applicability
of such calculations for breakup reactions was discussed before in Refs. 
\cite{bertsch,esben01}. In all reactions where the core survives (either
proton transfer or one-proton breakup) the matrix elements for the
transition $B\rightarrow A+p$ include the overlap integral $I_{Ap}^{B}(\vec{r%
})$ for the nuclei $A$, $p$, and $B$, obtained after integration over the
internal coordinates of $A$ for the many-body fully antisymmetric wave
functions, with $\vec{r}$ the vector connecting the center of mass of
nucleus $A$ with $p$ \cite{satch}. The ground states of the loosely bound
nuclei are known to be dominated by single particle features. Outside the
core both the nucleon-nucleon correlation effects and the antisymmetrization
effects are small and the overlap integral behaves very much like the radial
wave function for a single particle in the potential given by the core \cite
{blokh}:

\begin{equation}
I_{Aplj}^{B}(r)\rightarrow S_{nlj}^{1/2}\varphi _{nlj}(r)\rightarrow
C_{Aplj}^{B}\frac{W_{-\eta ,l+1/2}(2\kappa r)}{r}.  \label{asymp}
\end{equation}
Here $S_{nlj}$ is the spectroscopic factor and, in the rightmost part of the
equation, $C_{Aplj}^{B}$ is the asymptotic normalization coefficient
defining the amplitude of the tail of the overlap integral, $W$ is the
Whittaker function, $\kappa $ is the wave number, and $\eta $ is the
Sommerfeld parameter for the bound state ($Ap$). Due to Coulomb repulsion,
the radiative proton capture at stellar energies takes place at very large
distances from the core, in regions where approximation (\ref{asymp}) is
very good \cite{xu}, but the overlap integral there, and consequently the
resulting cross sections, are very small. However, the asymptotic
normalization coefficients $C_{Aplj}^{B}$ can be extracted from phenomena
that are peripheral, but involve regions of radii much closer to the core,
and therefore have larger cross sections and are easier to measure
experimentally. Another important reason for larger cross sections can be,
of course, the different nature of the transition operators involved.
One-proton removal reaction from loosely bound projectiles is one such
phenomenon. As an example to illustrate the above statement, for the well
known reaction $^{7}$Be(p,$\gamma $)$^{8}$B at solar energies (E$_{cm}$=20
keV) the classical turning point in the entrance channel is around $r\approx
250$ fm, but the combined effect of barrier penetration, transition operator
and asymptotic behavior of the $^{8}$B bound state wave function makes that
regions around  $r\approx 50$ fm contribute most in the radiative capture
process. This means that the wave function of $^{8}$B is most effectively
sampled at these distances, where it is very small. Going to processes where
distances $r\approx 4-5$ fm contribute most, like in the case of breakup
(fig.1 in Ref. \cite{tra01}), one can gain 5-6 orders of magnitude in cross
section from the magnitude of the wave function alone, as suggested
qualitatively above, while the only unknown quantity is the same ANC. We
treat the breakup reactions in this section using two approaches in a
Glauber model.

\subsection{Glauber model calculations with folded potentials}

In the extended Glauber model used here, the center of mass of the
projectile (made up of core and loosely bound proton) moves on a straight
line trajectory, an approximation valid at intermediate energies. The
relative motion of the outer proton about the core is described by a single
particle wave function with the asymptotic behavior given by Eq. (\ref{asymp}%
). They are assumed to interact separately with the target. The breakup of
the projectile appears from three different processes: the proton is
absorbed by the target while the remaining core is scattered and detected
(stripping); both the proton and the core are scattered by nuclear
interaction with the target and the core is detected (diffraction
dissociation); dissociation in the Coulomb field of the target. The
probability for each process depends on the impact parameter of each
trajectory and on the wave function for the proton-core relative motion. The
total cross section for this one-proton-removal reaction is given by the
integral over all impact parameters. S-matrix elements have been calculated
in the eikonal approximation including corrections up to second order \cite
{Wallace} to assure convergence. The S-matrix is given in this potential
approximation by 
\begin{equation}
S(b)=e^{i\chi (b)},  \label{eqr1}
\end{equation}
where the leading term in the eikonal expansion \cite{Wallace} is calculated
along the trajectory 
\begin{equation}
\chi (b)=-\frac{1}{\hbar v}\int_{-\infty }^{\infty }dzV(b,z).  \label{eqr2}
\end{equation}
The S-matrix elements are calculated for each trajectory, the contributions
to the three terms above are evaluated and then summed incoherently as 
\begin{equation}
\sigma _{sp}=\sigma _{str}+\sigma _{diff}+\sigma _{Coul}.  \label{sig-sp}
\end{equation}
For the reaction model calculations we assume that the ground state of the
projectile $(J^{\pi })$ can be approximated by a superposition of
configurations of the form $[I_{c}^{\pi _{c}}\otimes nlj]^{J^{\pi }}$, where 
$I_{c}^{\pi _{c}}$ denote the core states and $nlj$ are the quantum numbers
for the single particle wave function $\varphi _{nlj}(r)$\ in a spherical
mean field potential. These single particle states are normalized to unity
and have the asymptotic behavior given by Eq. \ref{asymp}, with the single
particle asymptotic normalization coefficients $b_{nlj}$. When more than one
configuration contributes to a given core state, then the total cross
section for one-nucleon breakup is written as an incoherent superposition of
single particle cross sections, weighted by the spectroscopic factors $%
S(c,nlj)$:

\begin{equation}
\sigma _{-1p}=\sum S(c,nlj)\sigma _{sp}(nlj).  \label{cs-tot}
\end{equation}
For $^{9}$C the last proton mainly occupies the $1p_{3/2}$ and $1p_{1/2}$
orbitals around a $^{8}$B core (which only has one bound state), in unknown
proportions. At large distances the two orbitals have identical radial
behavior and consequently have the same single particle cross section $%
\sigma _{sp}$ and the same single particle asymptotic normalization
coefficient $b_{p}$. The one-proton-removal cross section (\ref{cs-tot}) can
be written as the sum of two components: 
\begin{equation}
\sigma _{-1p}=\left[ S(1p_{3/2})+S(1p_{1/2})\right] \sigma
_{sp}(1p_{_{j}})=(C_{p_{3/2}}^{2}+C_{p_{1/2}}^{2})\sigma
_{sp}(1p_{j})/b_{p}^{2}  \label{SC2}
\end{equation}
and we can only extract the sum of the two spectroscopic factors or the sum
of the two (squares of) ANCs $C_{eff}^{2}=C_{p_{3/2}}^{2}+C_{p_{1/2}}^{2}$.
This same sum appears in the evaluation of the astrophysical $S$-factor.
Here we have only considered the $1p_{3/2}$ orbital. This was taken as the
solution of a radial Schr\"{o}dinger equation in a Woods-Saxon potential
with a geometry given by its half-radius and diffuseness parameters $(R,a)$
and a spin-orbit term with strength V$_{LS}$=18.6 MeV$\cdot $fm$^{2}$ \cite
{tra96}. The depth of the potential was adjusted to reproduce the
experimental separation energy of the last proton $S_{p}=1.296$ MeV, for
each geometry (different values of ($R,a$) used).

The coordinate system used in the Glauber calculations is shown in Fig. 1.
To calculate the S-matrix elements we need the interaction potentials
between the three participants. For the proton-target potential we have
used, as we did before, the $G$-matrix effective interaction of Jeukenne,
Lejeune and Mahaux (JLM) \cite{jlm}, in the updated version of Ref. \cite
{bauge}. This interaction is complex, energy and density dependent and was
adjusted for a very wide range of targets and proton energies. For the
target-core potential the double folding procedure described in \cite{tra-el}
was used: the same JLM interaction was folded with Hartree-Fock nuclear
matter distributions of the core and of the target. The single particle
densities were obtained in a spherical Hartree-Fock calculation using the
density functional method of Beiner and Lombard \cite{bei-lomb}. The
strength of the surface term in the functional was adjusted slightly in
order to reproduce the known experimental binding energy for each nucleus.
This procedure, similar to those used in Refs. \cite{bbs,brown01}, leads to
root-mean-square radii of the proton and mass distributions $%
<r_{p}^{2}>^{1/2}$=2.57 fm and $<r_{m}^{2}>^{1/2}$=2.43 fm, respectively,
for $^{8}$B, in fair agreement with our experimental determination using the
ANC method \cite{carb8}. The local density approximation was improved to
include finite range effects by using smearing normalized Gaussian functions
of ranges $t_{R}$ and $t_{I}$. The resulting double folding potentials were
subsequently renormalized to reproduce a variety of elastic scattering data
for light nuclei. We found in Ref. \cite{tra-el} that at incident energies
of about 10 MeV/u the real part of the potential needed a smearing with a
Gaussian of range $t_{R}=1.2$ fm and a substantial renormalization ($%
N_{V}=0.37$), while the imaginary part did not need renormalization ($%
N_{W}=1.00$) and the smearing was larger $t_{I}=1.75$ fm. In the present
calculations we adopted this procedure, with the JLM(1) interaction, and $%
N_{W}=1.00$. The procedure and its parameters determined by this approach
were found to give good predictions for the elastic scattering of
radioactive nuclei such as $^{7}$Be \cite{azhari3} and $^{11}$C \cite{tang}
at similar energies. We have also checked the procedure on a much wider set
of data from the literature at larger energies, and found a similar
conclusion for the imaginary part of the potential, while the
renormalization of the real part approaches unity around 50 MeV/u. The
S-matrix calculations that enter the first two terms of Eq. \ref{sig-sp}
depend primarily on the absorption, and thus on the geometry and strength of
the imaginary part of the interaction. The real part of the potential only
influences the phases of the scattered waves as the particles go along the
classical trajectories and does not affect the total cross section. The
Coulomb dissociation term is treated in a perturbative method equivalent to
that of Ref. \cite{bertulani}, except that radial matrix elements for $E1$
and $E2$ transitions were calculated with realistic Woods-Saxon radial wave
functions. Thus both nuclear and Coulomb breakup contributions were
calculated in a consistent way.

In order to check that the process was peripheral, the stripping and the
diffraction dissociation probabilities were calculated as a function of the
parameter $s$ (Fig. 1), which is the projection of the proton position
vector relative to the core, onto the plane perpendicular to the beam
direction. The stripping and the diffraction dissociation probability
distributions are shown in Fig. 2. It is clear that for all four targets
(the calculations were done for single isotopic targets $^{12}$C, $^{27}$Al, 
$^{116}$Sn and $^{208}$Pb) the contributions from both terms peak outside
the radius of the assumed $^{8}$B core, R$_{c}$=2.60 fm \cite{carb8}.
However, the interior does contribute and it should not be discarded. Note
that for all four targets the stripping cross section dominates, with the
diffraction dissociation part contributing only about 15\%. This situation
is different from that depicted in Fig. 1 of Ref. \cite{tra01} where the
same two probabilities are shown for the case of $^{8}$B breakup on a Si
target at 38 MeV/u. In that case the two components of the total cross
section have similar magnitudes, with the diffraction dissociation part
being slightly larger. This is a well understood bombarding energy effect.
High energy heavy ion interactions are dominated by strong absorption. The $%
^{8}$B calculations are in remarkable agreement with the experimental data
that explicitly disentangled the stripping and dissociation reaction
mechanisms \cite{nego}. The present results are for a much larger energy,
and the stripping term (absorption) dominates, as it does for $^{8}$B
breakup at larger energies. The third term that contributes to the total
one-proton-removal cross section, Coulomb dissociation ($\sigma _{Coul}$),
is the most peripheral due to the long range of the Coulomb forces. It was
calculated in its integral form and is not included in Fig. 2. In Fig. 9 of
Ref.\label{1} \cite{esben} its radial dependence is shown for situations
close to the present one. The present calculations were repeated, changing
only the half-radius and the diffuseness parameters of the proton binding
Woods-Saxon potential that describes the relative motion between the proton
and the core on a 20 point grid using $R=2.20-2.60$ fm in 0.10 fm steps and $%
a=0.50,$ $0.60,$ $0.65$ and $0.70$ fm, but keeping the reaction mechanism
parameters unchanged.

From the comparison of the calculations with the experimental cross sections
we extracted $C_{eff}^{2}=C_{p_{3/2}}^{2}+C_{p_{1/2}}^{2}$using Eq. \ref{SC2}%
. The results are compared in Fig. 3, in order, left to right, for the C,
Al, Sn and Pb targets. The error bars contain the contribution of the
experimental errors and of our uncertainties in the calculations, which are
described in detail below, added quadratically. The distribution of the four
numbers extracted is consistent with the constant value $C_{eff}^{2}=1.18\pm
0.12$ fm$^{-1}$. The results are also shown in the third column of Table I.

In a second approach, we used the free NN $t$-matrix interaction of Franey
and Love \cite{franey-love} in a $t\rho \rho $ approximation to obtain the
interaction potentials. This is a local representation of the free NN
interaction based on phenomenological nucleon-nucleon scattering amplitudes
of Arndt et al. \cite{arndt} at several energies between 50 and 1000 MeV. In
each spin-isospin channel the interaction is given by a linear combination
of Yukawa form factors. The longest range is fixed to be the long range part
of the one pion exchange potential ($OPEP$), but this part does not survive
in the direct term. In the calculation of nucleon-target and
projectile-target interactions, the free NN $t$-matrix should be
renormalized. We used the prescription suggested in \cite{franey-love}
employing relativistic kinematics. At intermediate energies one should
interpolate in the Franey-Love tabulation. In practice, this procedure is
difficult since the number of components varies from energy to energy. At
285 MeV/u we used the parametrization at the nearest energy (300 MeV). Other
effects such as Fermi averaging, Pauli blocking and spin-orbit contributions
were ignored. Finally, the scattering phase was calculated with Eq. (\ref
{eqr2}) in the lowest order of the eikonal approximation. We have checked
however that higher order noneikonal corrections are negligibly small at
this energy. The values extracted for $C_{eff}^{2}$ are similar with those
obtained with the JLM interaction and are presented in the forth column of
Table I.

\subsection{Glauber model calculations in the optical limit}

At 285 MeV/u, the optical limit of the Glauber model is also applicable. In
order to check the cross sections calculated above in the folding model, the
S-matrix elements were generated in the optical limit which is just the
leading term of the cumulant expansion of Glauber's multiple scattering
theory: 
\begin{equation}
\chi (b)=\frac{1}{2}\sigma _{NN}(i+\alpha _{NN})\int d\vec{b}_{1}d\vec{b}%
_{2}\rho _{proj}(b_{1})\rho _{targ}(b_{2})\tilde{v}(\vec{b}+\vec{b}_{1}-\vec{%
b}_{2})\text{,}  \label{chiNN}
\end{equation}
where $\rho _{proj(targ)}(b)$ is the profile density of the projectile
(target) obtained from the same Hartree-Fock densities used before: 
\begin{equation}  \label{eqg1}
\rho (b)=\int_{-\infty }^{+\infty }dz\rho _{HF}(b,z)  \label{rho}
\end{equation}
(z is the beam direction) and $\alpha _{NN}$ is the ratio of the real to the
imaginary nucleon-nucleon forward amplitude. These were interpolated for
each isospin state from Table I of Ray \cite{ray}. $\sigma _{NN}$ is the NN
total cross section at the given energy, taken from the parametrization of
Ref. \cite{s-john}. The integral appearing in Eq. (\ref{chiNN}) is a
projection onto the impact parameter plane of the interaction potential 
\begin{equation}
\Omega (R)=\int d\vec{r}_{1}d\vec{r}_{2}\rho _{proj}(r_{1})\rho
_{targ}(r_{2})\tilde{v}(\vec{r}_{1}+\vec{R}-\vec{r}_{2})  \label{eqf1}
\end{equation}
Following Ray \cite{ray} we identify the Fourier transform of the
interaction $\tilde{v}$ with the elementary scattering amplitude: 
\begin{equation}
\tilde{v}(q)=-\frac{2\pi \hbar ^{2}}{\mu }f=\frac{\hbar v}{2}\sigma
_{NN}(\alpha _{NN}+i)e^{-\beta _{NN}q^{2}}  \label{eqf2}
\end{equation}
where $q$ is the transferred momentum and $v$ is the asymptotic velocity. In
practical terms Eq. (\ref{eqf1}) was solved using a suitably chosen
elementary interaction $\tilde{v}$, such as 
\begin{equation}
\int d\vec{r}e^{i\vec{q}\vec{r}}\tilde{v}(r)=e^{-\beta q^{2}}  \label{eqf3}
\end{equation}
with the solution 
\begin{equation}
\tilde{v}(r)=\frac{1}{\pi ^{3/2}(4\beta )^{3/2}}e^{-\frac{r^{2}}{4\beta }}.
\label{eqf4}
\end{equation}
This was projected onto the impact parameter space and using Eq. (\ref{chiNN}%
) the scattering phase was calculated. The amplitudes $\alpha _{NN}$ and the
ranges $\mu =\sqrt{4\beta }$ were interpolated from Table I of Ray \cite{ray}%
. In the zero range approximation, $\mu \rightarrow 0$, Eq. (\ref{eqf1}) is
just the overlap volume of the two densities and the scattering phase is
determined by the elementary reactions in this volume. The breakup cross
section is highly sensitive to the range parameter. To assess this
sensitivity we have performed calculations with the ranges prescribed by
Ray, with zero-range for all interactions and then with an average value of $%
\mu =1.5$ fm in all $pp,$ $nn$ and $pn$ channels. This last approach is
denoted below as the ''standard'' procedure, as it is widely used to
estimate the size of halo nuclei from reaction cross sections \cite{bbs}.
When the incident projectile was a nucleon, then the corresponding profile
density was taken as a $\delta $ function. We use the charge independence
assumption i.e. $\sigma _{pp}=\sigma _{nn}$, $\sigma _{pn}=\sigma _{np}$, $%
\alpha _{pp}=\alpha _{nn}$, $\alpha _{pn}=\alpha _{np}$.

One test of the model and of the parameters used to calculate the breakup
cross sections, was to see how they reproduce the cross section data for the
assumed components of the $^{9}$C projectile ($p$ and $^{8}$B), separately,
on a $^{12}$C target (where the nuclear component dominates). We have
calculated the reaction cross section and the total cross section for$\ p$+$%
^{12}$C for energies E=100 to 1000 MeV, and the reaction cross section for $%
^{8}$B and for $^{9}$C in the energy range E/A=100 to 1000 MeV/u. The
results are compared with experimental data in Fig. 4. The proton reaction
cross section data (open circles) in the top panel are taken from the NNDC
data base \cite{NNDC} and those for the total cross section (full circles)
are from Ref. \cite{schw}. The reaction cross section data for $^{8}$B shown
in the middle panel are from Ref. \cite{blank} (filled circles) and Ref. 
\cite{tanihata} (open circle), while the reaction cross sections for $^{9}$C
in the lower panel are again from Ref. \cite{blank} (filled circle) and from
Ref. \cite{ozawa} (open circles), respectively. The calculations shown were
done in the zero-range approach, and give a good description of the data
over the whole range of energies. Therefore we have confidence that we can
proceed to breakup cross section calculations. We did the calculations with
the three range parametrizations described above: Ray, standard and
zero-range. For each parametrization, the calculations were repeated on the
grid of $(R,a)$ parameters for the proton binding potential, as described
above for the JLM\ potential calculations. Further details of all these
calculations will be presented elsewhere \cite{carst-glaub}. The results are
summarized in Table I and are shown in Fig. 5, where they are compared
directly with the values already shown in Fig. 3. The error bars are the
same as in the preceding figure, and the different points show the values
obtained using the three different prescriptions, as explained above. A good
consistence can be observed. We underline that the two Glauber approaches
are different in essence, and no further parameters were adjusted here. The
calculations done in the zero-range approximation gave smaller cross
sections (and consequently larger ANCs result) by 5\% to 16\% (less for Pb,
more for $^{12}$C target) and we exclude them from the overall average, as
the zero-range prescription is not very realistic. A similar check, done
with a larger range of the NN forces ($\mu =2.5$ fm) shows that the
extracted ANCs decrease by 10 to 25\%. These also are not included in the
average. These two extremes in the range parameter provide a lower and an
upper limit on the cross sections. The final average includes the results
obtained with four sets of calculations: JLM, Franey-Love, ''standard'' and
''Ray'' NN interaction. The weighted average is $C_{eff}^{2}=1.22\pm 0.13$ fm%
$^{-1}$, which agrees well with that found from the calculations in the
eikonal approach using the JLM interaction. The uncertainty takes into
account the experimental uncertainties, and the uncertainties in the
calculations. The ANC values were first averaged for each target, and then
an average was calculated for the four targets, using weights given by the
quadratically combined experimental and calculational uncertainties. To
evaluate the uncertainties for the calculated values we considered
independently those arising from the assumed nuclear structure (the geometry
of the proton binding potential), as described below, and those from the
reaction model. For each target we assumed that the standard deviation of
the values obtained from the four different calculations give a fair measure
of the reaction model contribution to uncertainty. Averaging first for each
type of calculations (as in the last row in Table I), and then for all
cases, leads to similar numbers ($C_{eff}^{2}=1.24\pm 0.13$ fm$^{-1}$), in
part because the experimental uncertainties dominate.

A discussion of the validity of the approach used here and of the
uncertainties occurring for the particular case of $^{9}$C at this energy, 
{\it a posteriori}, is in order. The use of such a fragile nucleus as $^{8}$%
B is as a core can be questioned, and also the high energy (285 MeV/u) may
lead to questions about the peripheral nature of the reaction. These
questions are supported by the non-negligible contributions of the interior
of the $^{9}$C nucleus, as shown in Fig. 2. The proton separation energy is
not as small here (1.296 MeV) as it was for $^{8}$B (0.137 MeV). Using the
procedure of Ref. \cite{carb8}, and the ANC extracted we find the rms radius
of the last proton in $^{9}$C to be $\langle r^{2}\rangle ^{1/2}=3.02$ fm,
considerably smaller than that found for $^{8}$B: $\langle r^{2}\rangle
^{1/2}=3.97$ fm. This, and the large separation energy, reflects in a less
sharp definition of the extracted ANC. In a figure equivalent to Fig. 2 of
Ref. \cite{tra01}, the ANC line has a larger slope, pointing toward the
conclusion that there is a residual dependence on the geometry of the proton
binding Woods-Saxon potential chosen. This contribution to the uncertainty
varies between 7.8\% and 4.6\% for the targets between C and Pb, and is
clearly smaller than the experimental uncertainties, and was considered in
the uncertainties plotted in Figs. 3 and 5. No new parameters were adjusted
in the present calculations. However some parameters were extracted before,
with corresponding uncertainties. For the case of the extended Glauber model
calculations with the JLM interaction, these parameters are the
renormalization ($N_{R}$, $N_{W}$) and the finite range ($t_{R}$, $t_{I}$)
parameters of the real and imaginary parts of the nucleus-nucleus potential
and have uncertainties. To see how these propagate to calculated breakup
cross sections, we repeated the calculations varying individually each
parameter. We discussed above that the real potential does not influence the
total cross section. Thus only the influence of $t_{I}$ and $N_{W}$ was
determined. We found the relative variation of the extracted ANC\ $\delta
C_{eff}^{2}/C_{eff}^{2}$ to be from $\pm $6.4\% on the C and Al targets,
down to $\pm $3.3\% on the Pb target (the relative contribution of the
nuclear breakup decreases as the charge of the target increases), when the
renormalization coefficient $N_{W}$ varies in the range found in Ref. \cite
{tra-el} $N_{W}=1.00\pm 0.09$. Similarly, a change of 4.7\% to 3.0\% was
found for a variation of 10\% around the central value of the smearing range 
$t_{I}=1.75$ fm. Combining the two in quadrature we obtain an 8\%
uncertainty, which is smaller than the experimental uncertainties are, and
similar to the uncertainties from the choice of the model. For the
calculations in the optical limit, the differences between the three
approaches used give an assessment of their overall uncertainty. The maximum
deviation around the average is less than 7-8\% in each case. Our present
model does not consider two-step processes like ${}^{9}$C$\to ^{8}$B$+p\to
{}^{7}$Be $+p+p$ and we cannot therefore directly assess their importance.
However, a recent paper \cite{khalili2000} compares Glauber model
calculations using a three-body projectile model with those using a two-body
projectile (as used in our calculations) and finds the dynamical
correlations to have a small influence (less than 10\%) on the calculated
one-neutron-removal cross section for the ($^{12}$Be,$^{11}$Be) and ($^{16}$%
C,$^{15}$C) reactions, a finding that supports the spectator assumption even
when the spectator itself is a loosely bound system, as in the case of $^{9}$%
C. The situations considered there are similar to the present one, as $^{11}$%
Be (assumed core for $^{12}$Be) and $^{15}$C (core for $^{16}$C) are halo
nuclei of an even larger extent than $^{8}$B (core for $^{9}$C) and,
therefore, expect the effect of two-step processes to be even smaller for
our case.

In a recent publication \cite{beaumel}, the ANC for $^{9}$C$\rightarrow ^{8}$%
B$+p$ was found using the proton transfer reaction $d$($^{8}$B,$^{9}$C)$n$
at 14.4 MeV/u incident energy. The experimental statistics were rather poor,
according to their Figure 1, and the authors present the results of a range
of DWBA calculations, with different optical potentials from literature.
They report an effective ANC that ranges from 0.97 to 1.42 fm$^{-1}$, with
an average that we found to be $\langle C_{eff}^{2}\rangle =1.18$ fm$^{-1}$.
From their assessment of the final uncertainty we figure $\delta
C_{eff}^{2}=0.34$ fm$^{-1}$. This uncertainty is large (30\%), and ($d,n$)
reactions have been criticized before \cite{gagl} for not being good
peripheral reactions, inadequate for the determination of the ANC. However,
their values (and particularly the average) are very close to those
extracted by us from different experimental data and using a different
reaction mechanism.

\section{The astrophysical factor $S_{18}$}

Since the ${}^{8}$B($p,$\thinspace $\gamma $)${}^{9}$C capture process at
astrophysical energies is a peripheral process, the absolute normalization
of its astrophysical $S_{18}$ factor is entirely defined by the ANC for $%
{}^{8}$B $+p$ $\to {}^{9}$C. To calculate the astrophysical $S$-factor we
used the potential model, as described in Ref. \cite{christy-duck}. Electric
dipole and quadrupole transitions were included for the final channel, with
E1 giving the largest contribution, and practically all waves were
considered in the entrance channel (but the $s$-wave dominates the major E1
term and the $d$-wave contributes only a few percent). The calculations were
done with a single proton $1p_{3/2}$ wave function normalized to unity and
having the asymptotic normalization coefficient $b_{p}$. Then the result was
scaled by $C_{eff}^{2}/b_{p}^{2}$ (such a procedure avoids any complications
that might appear when a Whittaker function normalized by $C_{eff}$ is used
over the whole integration range). The calculations were done for the proton
energy range $E_{cm}=0-0.8$ MeV. The contribution of the resonant state at $%
E_{res}=922$ keV with known width $\Gamma =100$ keV was found to be
unimportant here, because is rather far away and most likely its spin is $%
J^{\pi }=1/2^{-}$ and thus it is forbidden by selection rules to contribute
to the major term. A very weak dependence on energy is observed: $%
S(E)=45.8-15.1\cdot E+7.34\cdot E^{2}$ ($E$ in MeV), less than 15\% decrease
on the whole range.

Therefore, we find a value $S_{18}(0)=46\pm 6$ eV$\cdot $b$.$ This value is
in agreement with the value $S_{18}(0)=45\pm 13$ eV$\cdot $b reported in
Ref. \cite{beaumel}, but disagrees with those previously obtained from
various calculations. This radiative capture process was investigated in a
microscopic cluster model using the Generator Coordinate Method with two
different potentials (Volkov V2 and Minnesota) in Ref. \cite{desco}. The
author obtained for $S_{18}(0)$ values around 80 and 90 eV$\cdot $b,
respectively, therefore $80-100\%$ larger than our value. Earlier, Ref. \cite
{wies89} used a single particle model to estimate the S-factor and obtained
a much larger value (210 eV$\cdot $b), almost certainly because of the too
large spectroscopic factor $C^{2}S=2.5$ that they use. A short communication
by Timofeyuk \cite{nata} reported a value $S_{18}(0)=53$ eV$\cdot $b from
microscopic calculations using the M3Y nucleon-nucleon interaction, via an
estimate of the ANC and of the equivalent vertex constant, but without
giving too many details about the calculations. This last calculation is
closer to the experimental value determined here. Significant disagreement
between the astrophysical factors obtained in \cite{desco} and to a certain
extent in \cite{nata} using different potentials (although in slightly
different microscopic approaches) demonstrates the sensitivity of the
calculated ANCs to the effective NN-potentials adopted for the nuclear
structure calculations. Such sensitivity has been observed before for the 
{\it 1p}-shell nuclei in Ref. \cite{akr-tim}. Later the Generator Coordinate
Method (GCM) calculations with Volkov potentials \cite{desco94} led to an
ANC for the system ${}^{7}$Be $+p$ $\to {}^{8}$B about 70\% higher than the
ANC determined experimentally from proton transfer reactions \cite{azhari3}
and from the breakup reaction of ${}^{8}$B \cite{tra01}. The most
instructive dependence of the microscopically calculated ANCs on the adopted
effective NN-potential has been demonstrated by Baye and Timofeyuk (Ref. 
\cite{baye-tim}). The authors investigated the sensitivity of the nuclear
vertex constant (which is the ANC up to a trivial kinematical factor) for
the system ${}^{16}$O $+n$ $\to {}^{17}$O calculated in the microscopic GCM
for ten different effective NN-potentials. All the adopted potentials, among
which were 6 Volkov potentials, HNY, Minnesota and Gogny potentials,
significantly overestimated the experimental nuclear vertex constant, as
appears also to be the case here. For example, $V2$ (Volkov 2) potential
overestimates the ANC by 50\%. Hence the overestimation of the
microscopically calculated ANCs has a long history and the present result
for ${}^{8}$B $+p$ $\to {}^{9}$C once more seems to confirm that.

Using the astrophysical S-factor obtained above from the experimental ANC
for the energy region $E_{cm}=0-0.8$ MeV we evaluated the reaction rate
shown in Fig. 6 for the temperature range $T_{9}=0-1,$ which covers the
relevant temperature range for explosive hydrogen burning in supernovae \cite
{wies89}. Using the expansion for the reaction rate in powers of $T_{9}$ 
\cite{rolfs} (for the case of slowly varying astrophysical S-factors): 
\begin{equation}
R=N_{A}\langle \sigma v\rangle =T_{9}^{-\frac{2}{3}}\exp {\large (}-\frac{B}{%
T_{9}^{\frac{1}{3}}}{\Large )}(A_{0}+A_{1}T_{9}^{\frac{1}{3}}+A_{2}T_{9}^{%
\frac{2}{3}})\text{ }(\text{cm}^{3}/\text{s}/\text{mol})  \label{rate}
\end{equation}
where $N_{A}$ is Avogadro's number, $T_{9}$ is the temperature in 10$^{9}$
K, and 
\begin{equation}
B=3{\Large [(}\frac{\pi e^{2}Z_{1}Z_{2}}{\hbar }{\Large )}^{2}\frac{\mu }{2k}%
{\Large ]}^{1/3}=4.249[\mu {\rm [amu]}%
Z_{1}^{2}Z_{2}^{2}]^{1/3}~~(10^{9}K)^{-1/3}.  \label{b-temp}
\end{equation}
we found that we can reproduce the calculated temperature variation with the
constants: $B=11.945$ (not fitted), $A_{0}=6.64\cdot 10^{5}$, $%
A_{1}=8.50\cdot 10^{4}$ and $A_{2}=-2.41\cdot 10^{5}$. The formula fits the
integrated reaction rates with an accuracy of 0.5\% over the range $%
T_{9}=0-1.$

\section{Conclusions}

We have used existing experimental cross section data for the breakup of the
unstable nucleus $^{9}$C at 285 MeV/u on four different targets \cite{blank}
to determine the ANC of the radial overlap integral of the last proton
orbiting its core by comparing them with Glauber model calculations in the
eikonal approximation. Two different approaches were used to generate
S-matrix elements. In the first approach, both $G$-matrix and $t$-matrix
effective interactions were folded with HF single particle densities in
order to obtain the interaction potentials. In the second approach, the
optical limit of Glauber's multiple scattering theory was used with three
prescriptions for the elementary amplitudes. All calculations gave
consistent results. There were no new parameters that were adjusted in the
present calculations. We found the ANC for the virtual decay $^{9}$C$%
\rightarrow ^{8}$B$+p$, and then the astrophysical factor $S_{18}$ for the
radiative proton capture reaction $^{8}$B($p$,$\gamma $)$^{9}$C. This
reaction gives a possible path to the hot pp chain {\it pp-IV} at high
temperatures and away from it toward a rapid alpha process {\it rap I} at
high temperatures and densities \cite{wies89}.

Moreover, we show here a case where the indirect determination of nuclear
astrophysical information from the breakup of unstable nuclei at
intermediate energies proposed by us earlier for $^{8}$B, shows its
usefulness. The method we use was tested so far only for the case of $^{8}$B 
\cite{tra01} for which an independent knowledge of the ANC\ exists, and this
is a second case we propose. We also note that the breakup cross sections
are somewhat larger than those for the transfer reactions (around 100 mb,
compared with a few mb for proton transfer) due to the fact that essentially
the whole range of radii from about the grazing radius to infinity
participates, whereas for transfer only a limited region around the grazing
radius is involved.

Better experimental data, including detailed momentum distributions and
eventually disentangling the contribution of the stripping and diffraction
dissociation processes would provide additional information to check the
reliability of the models used. Also more breakup data in the same region of
nuclei would provide valuable information to check the parameters and
procedures used in these calculations and increase their overall reliability.

\section{Acknowledgments}

One of us (FC) acknowledges the support of IN2P3 for a stay at Laboratoire
de Physique Corpusculaire in Caen, during which part of this work was
completed. The work was supported by the U. S. Department of Energy under
Grant No. DE-FG03-93ER40773 and by the Romanian Ministry for Research and
Education under contract no 555/2000.

\begin{figure}[tbp]
\caption{ The coordinate system used in the Glauber model calculations. }
\label{fig1 }
\end{figure}

\begin{figure}[tbp]
\caption{ The stripping and diffraction dissociation contributions to the
breakup probability of 285 MeV/u $^{9}$C on C, Al, Sn and Pb targets as a
function of the proton impact parameter {\it s}. }
\label{fig2 }
\end{figure}

\begin{figure}[tbp]
\caption{ The asymptotic normalization coefficients determined using the
eikonal approach with potentials extracted using JLM interactions. The
results are plotted for the C, Al, Sn and Pb targets (left to right). The
uncertainties are a sum in quadrature of the experimental errors and the
uncertainties in the calculations. }
\label{fig3 }
\end{figure}

\begin{figure}[tbp]
\caption{ Comparison of the calculated reaction cross section (full line)
and total cross section (dashed line) for $p+^{12}$C (top panel), and of the
reaction cross section of $^8$B (middle panel) and $^9$C (bottom panel) on $%
^{12}$C target with experimental data. Data from Refs. \protect\cite
{blank,NNDC,schw,tanihata,ozawa} (see text). }
\label{fig7 }
\end{figure}

\begin{figure}[tbp]
\caption{ Same as Fig. 3, but the results of the calculations with
Franey-Love (squares), standard NN (triangles) and Ray (circles)
interactions are added to JLM (filled circles). }
\label{fig4 }
\end{figure}

\begin{figure}[tbp]
\caption{ The estimated reaction rate for the temperature range $T_9=0-1$. }
\label{fig6 }
\end{figure}

\newpage

\begin{center}
\begin{table}[tbp]
\caption{ Summary of the ANC extracted from different $^{9}$C breakup
reactions and five of the Glauber calculations. }
\label{tabl }
\end{table}
\begin{tabular}{|l|l|l|l|l|l|l|}
&  & JLM & Franey-Love & standard NN & Ray NN & zero-range NN \\ 
target & $\sigma (\delta \sigma )$ [mb] & $C_{eff}^{2}$ [fm$^{-1}$] & $%
C_{eff}^{2}$ [fm$^{-1}$] & $C_{eff}^{2}$ [fm$^{-1}$] & $C_{eff}^{2}$ [fm$%
^{-1}$] & $C_{eff}^{2}$ [fm$^{-1}$] \\ 
&  &  &  &  &  &  \\ 
C & 48(8) & 1.132 & 1.352 & 1.244 & 1.304 & 1.446 \\ 
Al & 55(11) & 1.069 & 1.167 & 1.077 & 1.125 & 1.220 \\ 
Sn & 146(31) & 1.482 & 1.448 & 1.391 & 1.438 & 1.496 \\ 
Pb & 181(40) & 1.183 & 1.140 & 1.115 & 1.145 & 1.172 \\ 
&  &  &  &  &  &  \\ 
average &  & 1.18 & 1.26 & 1.19 & 1.24 & 1.32
\end{tabular}
\end{center}

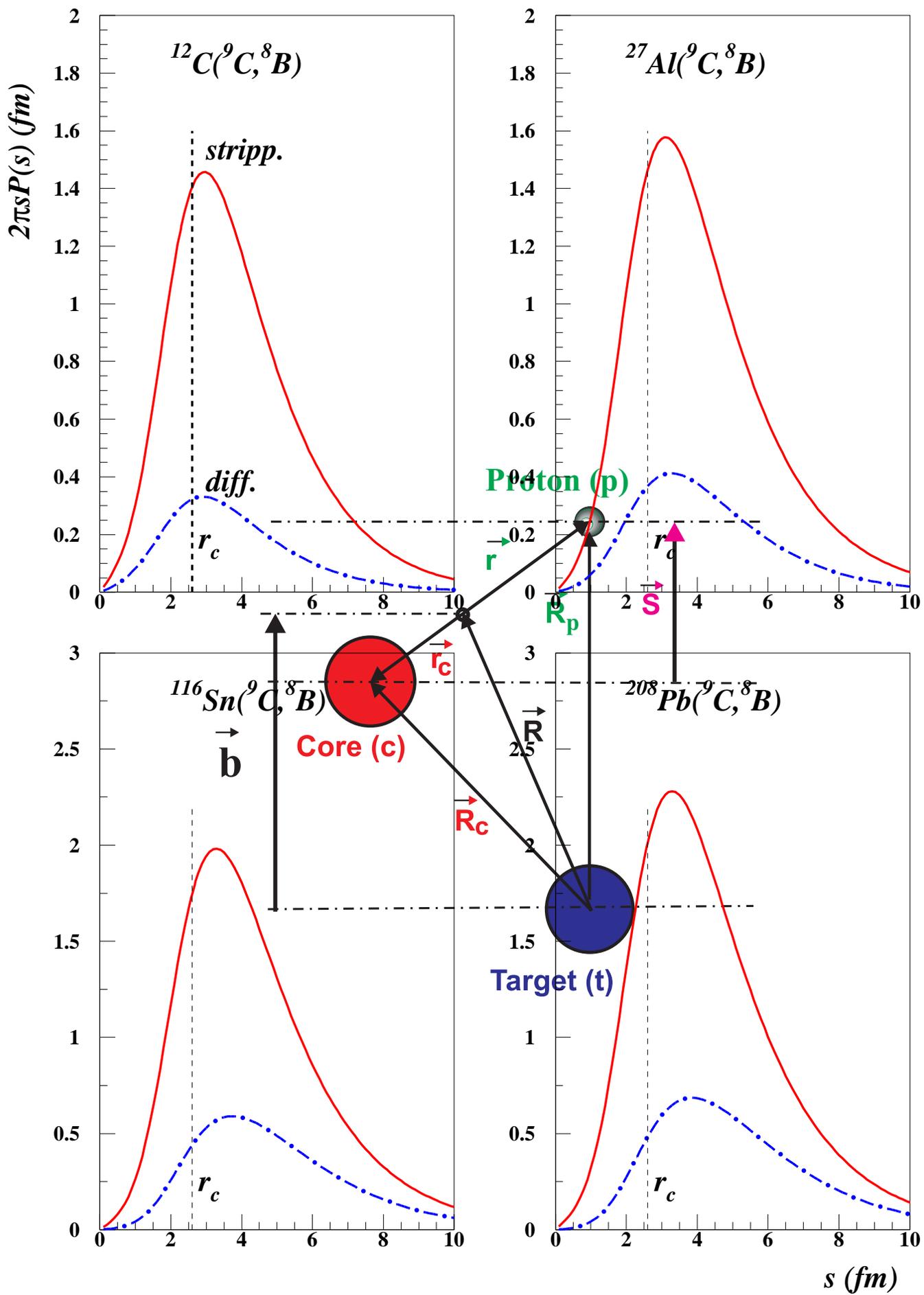
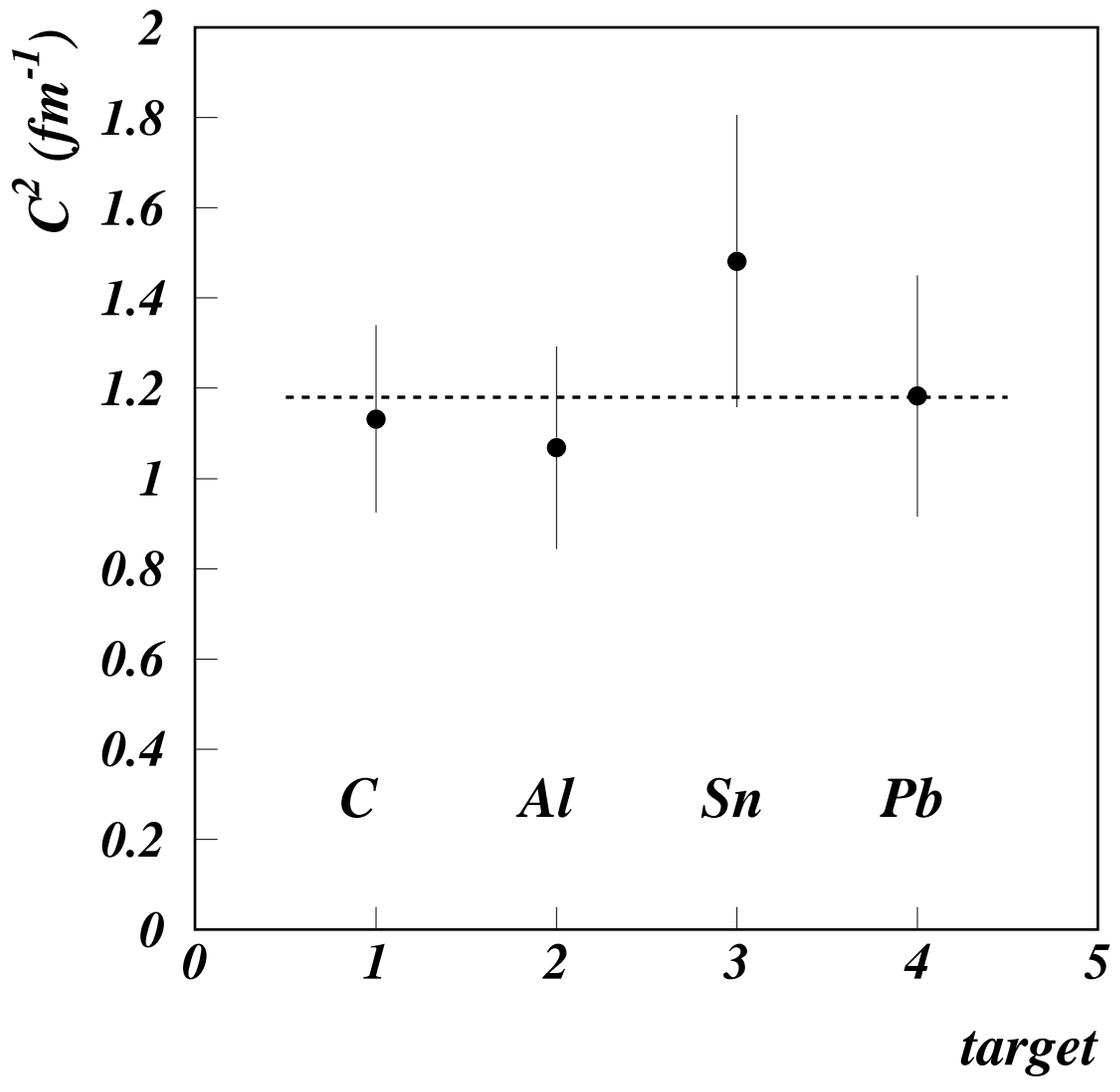
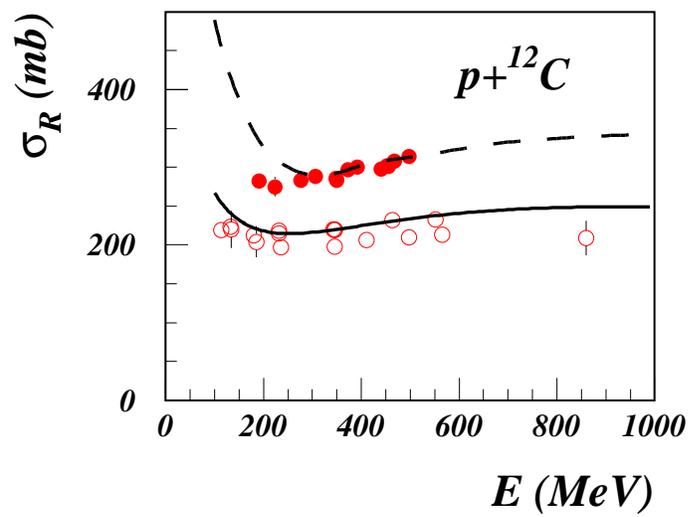
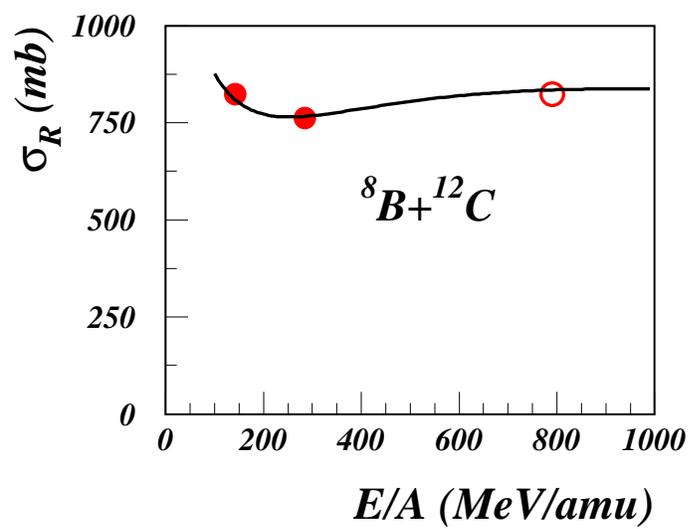
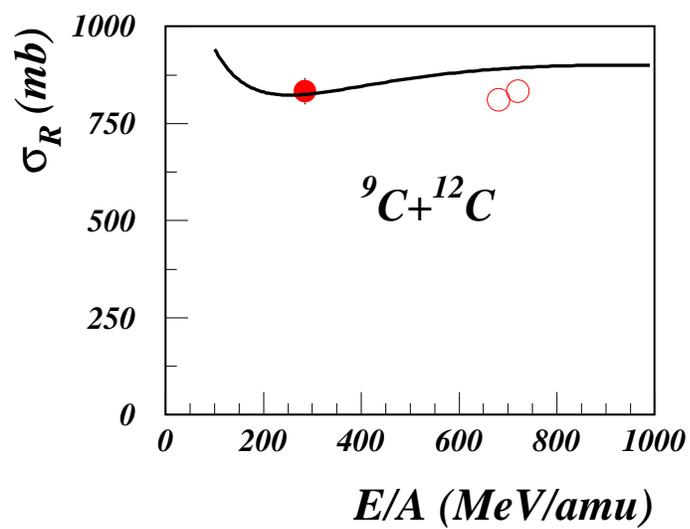
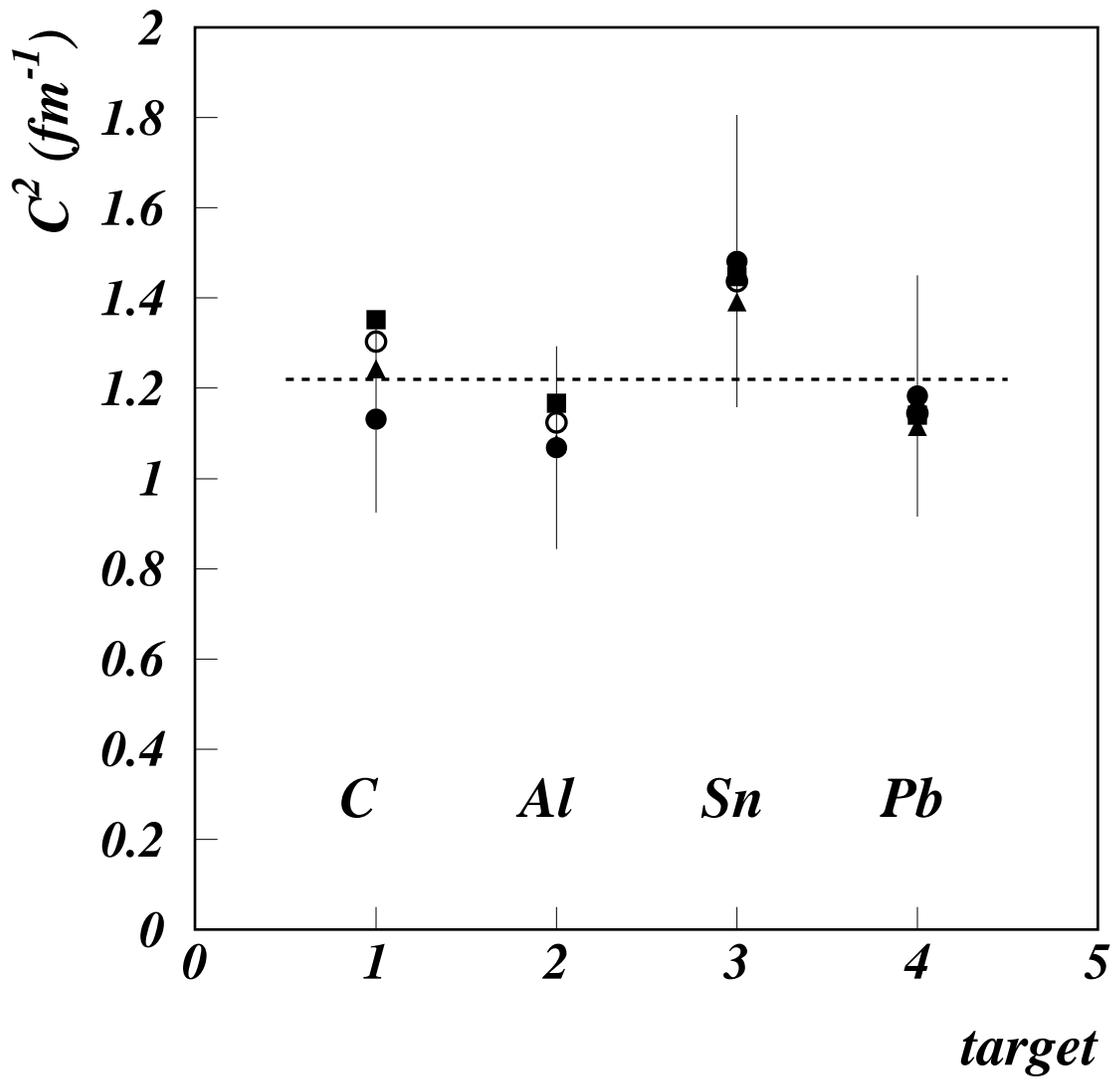
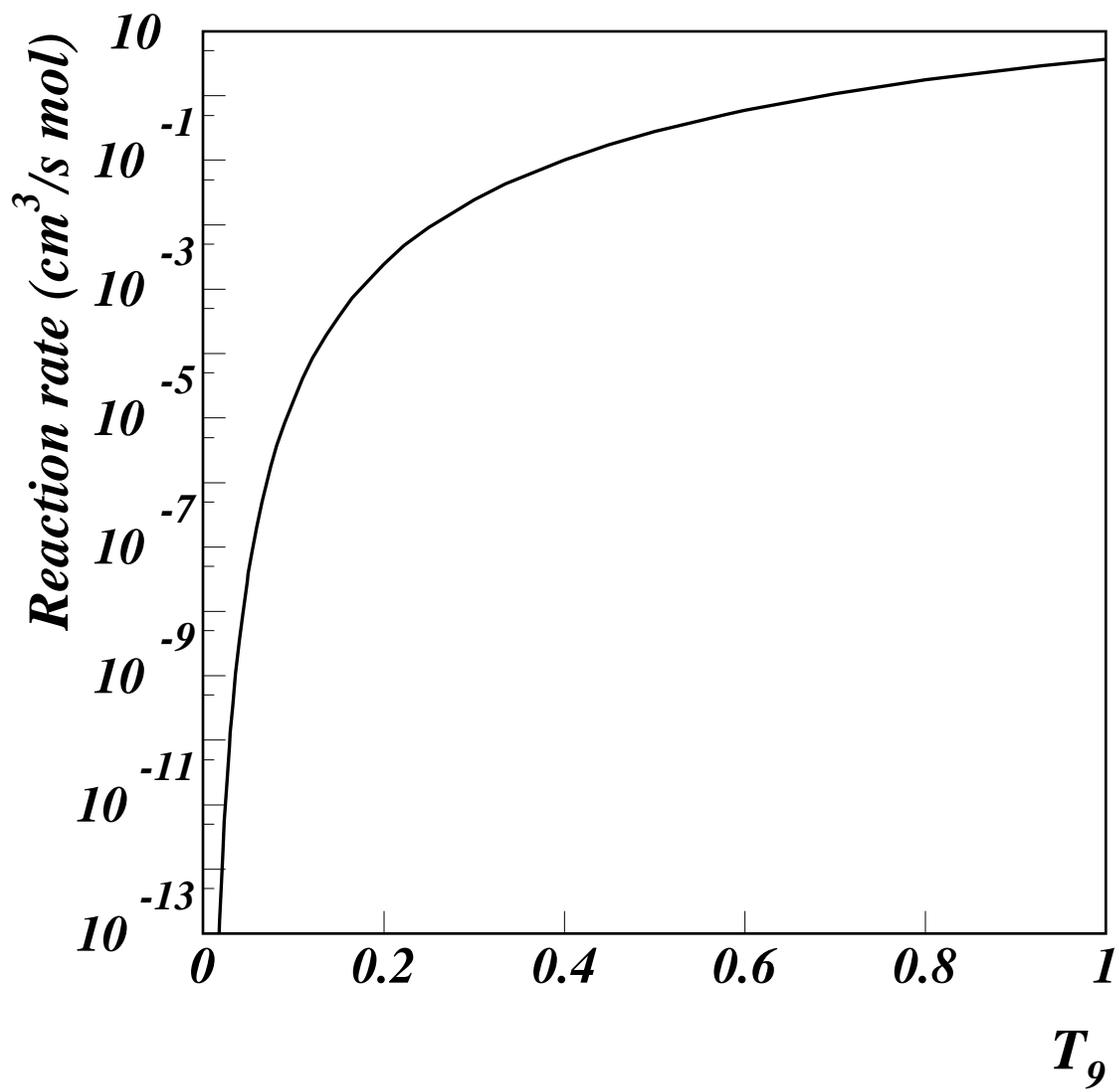
\end{document}